# Brillouin Scattering in Hybrid Optophononic Bragg Micropillar Resonators at 300 GHz

M. Esmann,[1] F. R. Lamberti,[1] A. Harouri,[1] L. Lanco,[1] I. Sagnes,[1] I. Favero,[2] G. Aubin,[1] C. Gomez-Carbonell,[1] A. Lemaître,[1] O. Krebs,[1] P. Senellart,[1] and N. D. Lanzillotti-Kimura,[1,*]

[1]Centre de Nanosciences et de Nanotechnologies, Centre National de la Recherche Scientifique, Université Paris-Sud, Université Paris-Saclay, Avenue de la Vauve, 91120 Palaiseau, France
[2]Matériaux et Phénomènes Quantiques, CNRS UMR 7162, Université Paris Diderot, 75013 Paris, France
*daniel.kimura@c2n.upsaclay.fr

*We introduce a monolithic Brillouin generator based on a semiconductor micropillar cavity embedding a high frequency nanoacoustic resonator operating in the hundreds of GHz range. The concept of two nested resonators allows an independent design of the ultrahigh frequency Brillouin spectrum and of the optical device. We develop an optical free-space technique to characterize spontaneous Brillouin scattering in this monolithic device and propose a measurement protocol that maximizes the Brillouin generation efficiency in the presence of optically induced thermal effects. The compact and versatile Brillouin generator studied here could be readily integrated into fibered and on-chip architectures.*

**Introduction**

Brillouin scattering,[1,2] the inelastic interaction of light with acoustic phonons, has been successfully exploited for a large number of applications. These range from novel approaches for all-optical data storage based on long-lived acoustic modes[3] to the tailored generation of optical frequency combs in periodically patterned semiconductor waveguides[4] and particularly narrowband Brillouin lasers in fibers[5] and semiconductors.[6] These applications require intense optical fields to enter the regime of stimulated Brillouin scattering and a good mode overlap between the optical and the acoustic field for the scattering process to be efficient. Therefore, many realizations of Brillouin scattering generators rely on optical fibers[7–9] or photonic waveguides[6,10,11] providing lateral confinement on the micrometer scale over a large interaction length in the centimeter range. Much effort has been devoted to the generation or mitigation of tailored Brillouin spectra in these devices either by introducing periodic patterns along the fiber using for instance the concept of distributed Bragg reflectors (DBRs)[12–14] or by specialized cross-sections in photonic crystal fibers[15,16] and suspended waveguides[10]. With these approaches, acoustic phonons up to few tens of GHz can be effectively controlled, a limit imposed by the minimum achievable feature size, which needs to be on the length scale of the phonon wavelength. In addition, these approaches to engineering the phonon spectrum have to take into account the simultaneous modifications introduced to the optical dispersion relation of the waveguides[10], which imposes some constraints on their versatility.

A very different class of devices that are employed for the generation and enhancement of Raman and Brillouin signals are resonant microstructures such as silica microspheres[17] or planar Fabry-Perot semiconductor microcavities[18–21] grown by molecular beam epitaxy. In contrast to fibers and waveguides, a large effective interaction length is achieved through the optical cavity Q-factor. More importantly, semiconductor planar microcavities have been used to embed ultra-high frequency multilayer nanophononic resonators confining acoustic phonon modes in the 100 GHz – 1 THz range[18,22–26]. This approach allows an independent design of the acoustic and optical densities of states. The acoustic multilayer only acts as an effective medium for the optical fields. However, these structures are by definition extended in the lateral dimension.

Here, we transfer the concept of nested, independently tunable opto-phononic resonators to three-dimensional semiconductor micropillars[20,21,27–29]. We thus combine the advantages of resonant cavities and transversal optical field confinement in a fiber, yet with minimum vertical feature sizes well below 10 nm. We develop an optical free-space technique to characterize spontaneous Brillouin scattering in such a monolithic device hosting a nanophononic interface mode at 320 GHz. We furthermore demonstrate a measurement protocol that allows us to maximize the Brillouin generation efficiency in the presence of optically induced thermal effects. Compact and versatile opto-phononic micropillar resonators could be integrated with existing fibered and on-chip architectures, opening a wide range of future applications.

**Brillouin spectroscopy on micropillars**

The sample under study is grown on a (001)-oriented GaAs substrate by molecular beam epitaxy (MBE). It consists of an optical microcavity with two distributed Bragg reflectors (DBRs) enclosing a resonant spacer with an optical path length of $2.5\lambda$ at a resonance wavelength around $\lambda = 890\text{nm}$. The top (bottom) optical DBR is formed by 14(18) periods of $Ga_{0.9}Al_{0.1}As/Ga_{0.05}Al_{0.95}As$ bilayers optimized to confine an optical mode with typical Q-factors of 2000. The optical spacer of the cavity is composed of two concatenated acoustic superlattices (SLs) with top (bottom) acoustic superlattices formed by 16 periods of 7.3nm/9.8nm (8.5nm/8.2nm) GaAs/AlAs layers. This acoustic structure is designed to confine acoustic phonons at 300GHz at the interface between the two SLs based on their different topological properties[30–33]. From this nested planar opto-phononic cavity, 30µm pitched arrays of square and circular micropillars with various lateral sizes are fabricated by optical lithography and inductively coupled plasma etching. Scanning electron microscope (SEM) images of an array of micropillar resonators and a zoomed-in view of a single square micropillar with 4.5µm lateral extent are shown in Figure 1a and b, respectively. We numerically calculate the electric field distribution of the fundamental optical mode in a circular micropillar resonator through the Finite Element Method (FEM) as shown in Figure 1c. In the optical domain, the high frequency acoustic resonator behaves as an effectively homogeneous medium with optical properties modulated on a deep sub-wavelength scale. The micropillar thus behaves as a three-dimensional resonator with wavelength-scale optical confinement in all three dimensions of space, resulting in a discrete spectrum of optical modes. In the vertical direction, the resonant cavity confinement leads to an exponentially decaying envelope of the optical mode in both DBRs. In the radial direction, the refractive index contrast between the semiconductor materials and vacuum leads to an additional in-plane confinement of the mode with a Bessel-type envelope[34–36] much like for an optical fiber.

To probe the Brillouin spectrum of individual micropillars, an incident laser beam (mode hop-free cw Ti:Sa laser M2 SolsTiS) resonant with the fundamental optical micropillar mode is focused to a spot size of roughly 10µm (FWHM) on the sample surface and centered on a pillar. Excitation and collection are realized through the same optical elements under normal incidence. The main obstacle in Brillouin scattering measurements on wavelength-scale objects is stray light rejection. Usually, there is a compromise between the minimum size of the studied object (source of the stray light) and the minimum Brillouin frequency that can be accessed stray light-free. We overcome this experimental challenge by actively using the spatial mode mismatch between the incoming laser beam and the optical micropillar modes as a signal filtering method, as sketched in Figure 1d. Three contributions emerge from the sample: the reflection of the excitation laser from the substrate surface (red arrows), the reflection of laser light from the micropillar, and the Brillouin signal. In contrast to the other two, the Brillouin signal originates solely from the fraction of laser light coupled into the micropillar and interacting with the nanoacoustic structure constituting the resonant optical spacer. It hence emerges from the micropillar through its optical modes with a spatial pattern approximated by a Gaussian beam (blue). In a plane behind the collection lens, the directly reflected contributions lead to the formation of a pronounced diffraction pattern. The experimentally observed pattern is shown in the right part of Figure 1d. We optimize the selective collection of the Brillouin signal by inserting a spatial filter behind the collection lens, exclusively transmitting in a node of the diffraction pattern where the contribution emerging from the cavity dominates. Figure 2a presents the Brillouin signal obtained using the spatial filtering method described above recorded on a 4.5 μm square micropillar. It exhibits three pronounced peaks at acoustic frequencies around 280GHz, 320GHz, and 360GHz, labelled A, B and C.

Here, we note that the acoustic characterization of semiconductor micropillars working at 20GHz has been previously performed by means of impulsive coherent phonon generation and detection.[20,37] While a powerful characterization tool, it is however not applicable for the investigation of spontaneous Brillouin scattering. With standard Brillouin or Raman scattering techniques, the frequency range explored in this work is usually inaccessible in micro-objects due to insufficient stray-light rejection. Also standard radio frequency noise spectroscopy often used with optomechanical microresonators is usually limited to a few GHz. The present observations thus represent the first Brillouin scattering measurement on a micron sized acoustic resonator in the 100GHz frequency range.

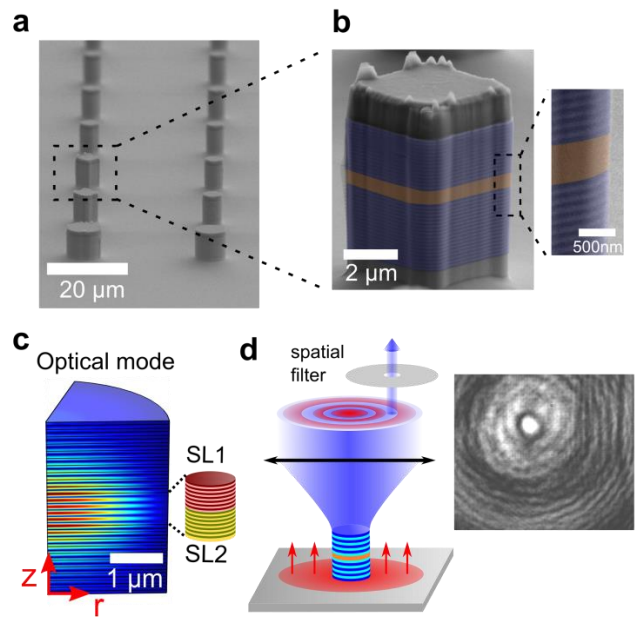

*Fig. 1. **(a)** Scanning-electron micrograph (SEM) of an array of circular and square micropillar resonators and **(b)** of a single square micropillar with 4.5 μm lateral extent. The top layer is SiN deposited as part of the dry etching. **(c)** Finite element method (FEM) simulation of the fundamental optical mode in a circular micropillar, showing the absolute value of the electric field. The resonant optical spacer is composed of two nanoacoustic superlattices (SL). **(d)** The reflected optical signal comprises direct reflection from the substrate, reflections from the micropillar and the Brillouin spectrum (blue) with a spatial pattern dictated by the optical micropillar modes. The reflected contributions form a pronounced diffraction pattern (red rings, experimental pattern in the inset). A spatial filter (grey disk) optimizes the relative collection ratio of the Brillouin signal.*

To understand the origin and nature of the observed peaks, Brillouin spectra are also measured on the planar microcavity, not etched into pillar shapes. In this case, a perfect spatial filtering of the reflected light can be performed by simply blocking the reflected beam in the collection path. By furthermore taking advantage of the angular dispersion of the planar microcavity, a simultaneous incoming-outgoing optical resonance condition can be established[30,38,39]. A Brillouin spectrum on the planar cavity is shown in Figure 2b showing very similar peaks as for the micropillar. In the case of a planar structure, the corresponding high-frequency acoustic modes can be modelled by means of transfer-matrix calculations[40]. The calculation results shown in Figure 2c confirm that the strongest, central peak in the Brillouin spectrum is caused by the topological phononic interface state that presents the largest overlap with the cavity optical mode. In contrast, the two other peaks correspond to propagating modes extending over the full acoustic structure. These latter modes are a general feature in Brillouin scattering of periodic superlattices[41,42] and fulfill the backscattering condition $k_p = 2k_l$ with $k_p$ the (quasi-) momentum of the phonon and $k_l$ the momentum of the laser. To fully account for the shape and relative composition of the observed Brillouin spectrum, we furthermore perform calculations of the anti-Stokes Brillouin scattering cross-section based on a 1D photoelastic model (dashed red in Figure 2b)[42]. Note that at room temperature the fundamental electronic transition in GaAs is at 1.52 eV, such that the experiments are performed under nearly-resonant conditions. The Brillouin spectrum is thus dominated by the photoelastic contributions from the GaAs layers. The results of the calculation shown in Figure 2b well reproduce the measured Brillouin spectrum. The larger peak width in the experimental spectra is a consequence of finite spectral resolution in our setup of approximately 13 GHz[30].

The close similarity between the Brillouin spectra of the micropillar and the planar cavity is a direct consequence of the high acoustic resonance frequencies around 300 GHz explored here. These frequencies correspond to acoustic wavelengths around 10 nm, which are roughly one order of magnitude smaller than the optical wavelengths involved in the Brillouin scattering process. In the acoustic domain, the micropillar therefore confines longitudinal acoustic phonons to wavelength scale dimension along the vertical direction. However, since the characteristic wavelength of the confined mechanical mode is much smaller than the micropillar's lateral size, the system effectively behaves as an infinitely extended planar acoustic structure in the lateral direction.

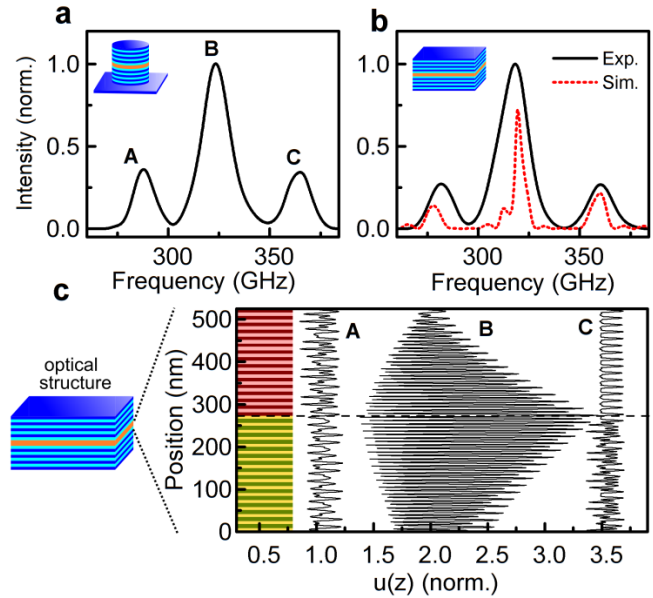

*Fig. 2. (a) Anti-stokes Brillouin spectrum measured on the micropillar resonator shown in Figure 1b. The spectrum exhibits three pronounced peaks (A-C). (b) Brillouin spectrum measured on a planar resonator with identical vertical structure (solid black). A photoelastic model calculation (dashed red) well accounts for the three-peaked structure of the spectrum. (c) Simulated phonon modes corresponding to the three main peaks in panels (a) and (b). Displayed is the absolute value of the mechanical displacement $u(z)$.*

**Optimizing Brillouin scattering in the presence of thermal effects**

While Brillouin spectra in the planar and micropillar resonators studied here are spectrally very similar, striking differences are observed in their dependence on laser power. We recorded optical reflectivity spectra as a function of power on the micropillar resonator shown in Fig. 1. The results for laser powers between 1mW and 34mW incident on the focusing lens are displayed in Fig. 3a. For 1mW, we find a symmetric reflectivity dip centered at 892.3 nm with a minimum of 65% in normalized reflectivity and a Q-factor of 2000. The contrast is mainly limited by the mode mismatch between the excitation laser and the fundamental micropillar mode[43] (see Fig. 1). With increasing laser power, the reflectivity minimum shifts to larger wavelengths reaching a maximum shift of

$\Delta\lambda_{res} = 0.4\text{nm}$ (see panel c). Furthermore, the shape of the reflectivity becomes asymmetric with a steeper slope at the long wavelength end. Note that in all curves the laser was scanned from short to long wavelength. For comparison, we performed an equivalent experiment on the planar, not etched portion of the microcavity (gray crosses in panel c). Here, the power-dependent shift in resonance wavelength is almost absent. These observations are attributed to power dependent thermal effects in micropillars. Indeed, working relatively close to the band edge in GaAs at room temperature, heating and subsequent modification of the electronic band structure of the materials are expected with increasing excitation power. While in the case of a planar microcavity heat can be rapidly dissipated both through the substrate and in the lateral direction, the finite lateral size of a micropillar implies a limited conduction of heat towards the substrate, and a much more pronounced light-induced rise in temperature.

With increasing temperature, the refractive indices of GaAs and AlAs rise[44], resulting in an increase of the optical path length through the cavity and hence an effective red-shift of the resonance wavelength. For each power and wavelength, the system acquires a stationary equilibrium for which the absorbed optical power and dissipated heat flow into the substrate are of equal magnitude. We describe this self-consistent problem in terms of a Lorentzian resonator with power-dependent resonance energy [28] based on the parameters deduced from the experimental reflectivity curves in Figure 3a and 3c. Calculation results are displayed in Figure 3b and 3d. We observe that a simple one-resonator model can consistently reproduce both the shapes of the reflectivity spectra and power-dependent resonance shifts.

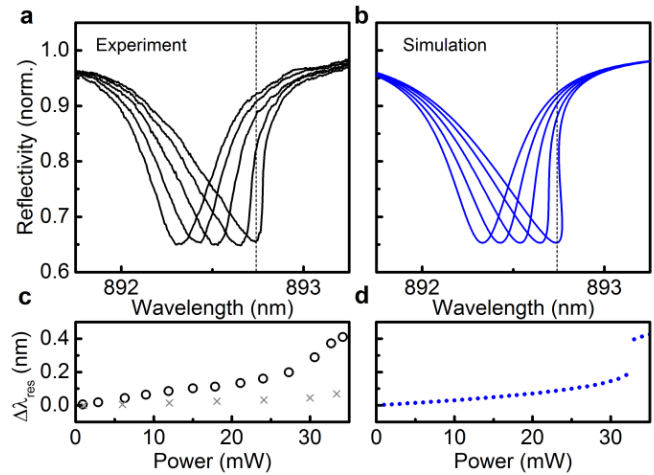

*Fig. 3. (a) Optical reflectivity of the micropillar resonator shown in Fig.1. recorded at laser powers of 1mW, 9mW, 18mW, 27mW and 34mW. With increasing laser power absorption-induced heating leads to a systematic red-shift of the optical resonance. (c) For an excitation wavelength of 892.7nm (dashed line in panel a) an overall power-dependent red-shift of 0.4nm is found. On a planar cavity with the same vertical structure, a shift of only 0.05nm is found under equal excitation conditions (gray crosses). (b) Simulated reflectivity spectra based on a self-consistent one oscillator model[28]. (d) Simulated power-dependence of the cavity resonance extracted from the modelling results in panel b.*

These drastic changes in the optical device properties due to laser-induced heating will affect its operation as a Brillouin generator. Indeed, the Brillouin scattering cross-section strongly depends on the coupling between the excitation laser and the optical cavity mode. Building upon the results presented in Fig. 3, we measure power-dependent Brillouin spectra on both the micropillar resonator and a planar microcavity with identical vertical structure. For the micropillar, we tune the laser to 892.7nm (dashed vertical lines in Fig. 3). That is, the wavelength is chosen such that the excitation is off-resonant at low power, but becomes on-resonant at high power through optical heating. The quantity plotted in Figure 4a is the integrated area under the central Brillouin peak (see upper inset) normalized to incident laser power. We furthermore set the area measured at a power of 1 mW to unity. Therefore, a linear power dependence of the spontaneous Brillouin signal is represented by a constant value of unity. However, for the micropillar (black circles) we observe an increase in Brillouin efficiency by a factor of 25. For comparison, we perform the equivalent measurement on a planar cavity (gray crosses) resulting in an almost power-independent normalized Brillouin efficiency around unity. These observations arise from the absorption-induced temperature changes discussed above and the related red-shift of the optical micropillar resonance. We perform corresponding photoelastic model calculations using the experimentally determined power-dependent shift in optical resonance wavelength (Fig. 3c) to calculate power-dependent laser- and scattered fields. The calculated Brillouin spectra are then processed in exactly the same manner as their experimental counterparts. To reach the quantitative agreement shown in Figure 4b, we furthermore make the assumption that the laser field and the scattered Brillouin field are equally enhanced upon the power-dependent red-shift of the micropillar mode. Although the measured enhancement in Brillouin signal is very reminiscent of the onset of stimulated Brillouin scattering, our modeling conclusively demonstrates that thermal effects on the optical cavity mode fully accounts for our experimental observations.

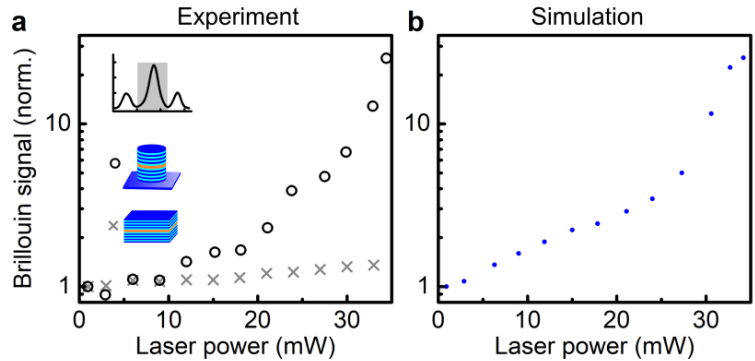

*Fig. 4. (a) Power-dependent Brillouin spectra are recorded on the micropillar shown in Fig. 1 (circles) with the excitation laser red-detuned from optical resonance by 0.4nm at 1mW of power (dashed vertical line in Fig. 3a). Plotted is the area under the central Brillouin peak (upper inset) normalized to laser power. The area measured at 1mW is set to unity. The normalized signal increases 25-fold when increasing the laser power to 35mW. On a planar cavity with the same vertical structure (grey crosses) the increase in normalized signal is almost absent. (b) Photoelastic model calculation based on the structure in Fig. 2 and experimentally measured power-dependent resonance shifts (Fig. 3c). Resonant optical excitation and collection are assumed.*

**Discussion and Conclusions**

The present results demonstrate that optical micropillar cavities are an interesting platform for the generation of high frequency Brillouin spectra. An apparent difficulty of using a micron-sized platform arises from the inevitable generation of stray light, which we overcome here by proposing a spatial filtering technique. This allows lifting the usual tradeoff between object size and minimum accessible frequency in the Brillouin spectrum. The proposed structure also offers great versatility since any planar multilayer structure that shapes the acoustic phonon spectrum such as mirrors, filters, resonators and coupled cavity systems [18,45–53] could be embedded in the optical cavity if its characteristic feature size is small enough to act as an effective optical medium, i.e. working at ultra-high phononic frequencies.[18,25,30,54] We have furthermore shown that one can exploit heating effects induced by the excitation laser to lock the optical cavity mode at a given spectral position for a given power and to optimize the Brillouin scattering signal by initially detuning the excitation laser from the unperturbed cavity mode.

Building upon these results, several possible routes could be followed to enter the stimulated Brillouin regime: Since the monolithic cavity studied here is based on DBRs, the optical quality factor could be enhanced by orders of magnitude by increasing the number of DBR layers to access Brillouin lasing.[6] Likewise, the acoustic quality factor may be raised to target phonon lasing.[55] Moving the resonance of the optical cavity closer to the electronic band gap in the materials of the nanoacoustic structure the photoelastic interaction can be increased resulting in resonant Brillouin scattering.[19] Finally, the compactness and waveguiding structure allow integration of our device with existing fibered and on-chip architectures.[56,57]

**Acknowledgements**

The authors acknowledge funding by the European Research Council Starting Grant No. 715939, Nanophennec; by the French RENATECH network, by the French ANR (QDOM project) and through a public grant overseen by the ANR as part of the "Investissements d'Avenir" program (Labex NanoSaclay Grant No. ANR-10-LABX-0035). ME acknowledges funding by the Deutsche Forschungsgemeinschaft (DFG, German Research Foundation) – Project 401390650.